\documentclass[useAMS,usenatbib]{mn2e}
\usepackage{psfig}

\title[Halo assembly bias and its effects on galaxy clustering]
{Halo assembly bias and its effects on galaxy clustering}

\author[D.~J.~Croton, L.~Gao \& S.~D.~M.~White]{
\parbox[t]{\textwidth}{
Darren J. Croton$^1$,
Liang Gao$^2$,
Simon D. M. White$^3$
}
\vspace*{6pt} \\ 
$^1$Department of Astronomy, University of California, Berkeley, CA, 94720, USA \\
$^2$Institute for Computational Cosmology, Physics Department, Durham. U.K.\\
$^3$Max-Planck-Institut f\"ur Astrophysik, D-85740 Garching, Germany
\vspace{-0.5cm} 
}

\date{Accepted ---. Received ---;in original form ---}

\pubyear{2005}

\newcommand{\plotone}[1]
           {\centering \leavevmode \psfig{file=#1,width=\columnwidth,clip=}}

\newcommand{\plotfull}[1]
           {\centering \leavevmode \psfig{file=#1,width=15cm,clip=}}

\def\simlt{\lower.5ex\hbox{$\; \buildrel < \over \sim \;$}}
\def\simgt{\lower.5ex\hbox{$\; \buildrel > \over \sim \;$}}

\begin{document}

\maketitle


\begin{abstract} 
The clustering of dark halos depends not only on their mass but also
on their assembly history, a dependence we term `assembly bias'. Using
a galaxy formation model grafted onto the Millennium Simulation of the
$\Lambda$CDM cosmogony, we study how assembly bias affects {\it
galaxy} clustering.  We compare the original simulation to `shuffled'
versions where the galaxy populations are randomly swapped among halos
of similar mass, thus isolating the effects of correlations between
assembly history and environment at fixed mass. Such correlations are
ignored in the halo occupation distribution models often used populate
dark matter simulations with galaxies, but they are significant in our
more realistic simulation. Assembly bias enhances 2-point correlations
by 10\% for galaxies with $M_{\rm b_J}\!-5\log h$ brighter than $-17$,
but suppresses them by a similar amount for galaxies brighter than
$-20$.  When such samples are split by colour, assembly bias is 5\%
stronger for red galaxies and 5\% weaker for blue ones.  Halo central
galaxies are differently affected by assembly bias than are galaxies
of all types. It almost doubles the correlation amplitude for faint
red central galaxies. Shuffling galaxies among halos of fixed
formation redshift or concentration in addition to fixed mass produces
biases which are not much smaller than when mass alone is fixed.
Assembly bias must reflect a correlation of environment with aspects
of halo assembly which are not encoded in either of these parameters.
It induces effects which could compromise precision measurements of
cosmological parameters from large galaxy surveys.
\end{abstract}

\begin{keywords}
cosmology: theory, galaxies: evolution, galaxies: clustering
\end{keywords}

\section{Introduction}
\label{intro}

In a recent study \citet[][hereafter GSW05]{Gao2005} showed that the
clustering of dark matter halos can depend strongly on their formation
redshift.  Many current galaxy clustering models adopt simplified
prescriptions for populating halos with galaxies based on an implicit
assumption which is inconsistent with this result, namely that the
assembly history of a halo of given mass (and thus its galaxy content)
is statistically independent of its larger scale environment
\citep[e.g.][]{Kauffmann1997,Jing1998,Peacock2000, Benson2000,
Berlind2003, Yang2003}. GSW05 found that those halos with $M_{\rm
vir}\!<\!10^{12}{\rm M}_\odot$ that assembled at high redshift are
substantially more clustered than halos of similar mass that assembled
more recently.  Earlier studies had missed the strength of this
dependence \citep[e.g.][]{Lemson1999,Sheth2004}, apparently because
they were based on simulations of insufficient size and resolution to
reliably reach the relevant regime. Following GSW05, other authors
have demonstrated significant dependences of clustering on halo
properties such as concentration and subhalo occupation number, which
are strongly correlated with formation redshift
\citep[e.g.][]{Wechsler2006, Zhu2006}.

If the assembly history of dark matter halos is correlated with their
large-scale environment, we may expect the same to be true for their
\emph{galaxy} content. This will then affect the large-scale
clustering of galaxies in a way which depends on how galaxy properties
are established during halo assembly, i.e. on the physics of galaxy
formation.  A number of recent studies have addressed this question,
approaching it from both observational and theoretical points of view
\citep[e.g.][]{Yoo2006,Harker2006, Yang2006, Abbas2006,
Reed2006}. \cite{Yang2006} find that, at fixed mass, the clustering of
galaxy groups correlates quite strongly with the star formation rate
of the central galaxy. On the other hand, \cite{Skibba2006} and
\cite{Abbas2006} found the clustering in their analysis of SDSS data
to be consistent with models with no dependence of halo galaxy
populations on halo environment.  \cite{Yoo2006} randomly shuffle
galaxies between halos of similar mass in a small volume simulation
(box side length $50\,h^{-1}\rm{Mpc}$) and find $5-10\%$ effects that
are at the level of the statistical uncertainty of their
calculation. Much of this work was a response to the presentation of
early results from the present project during summer
2005\footnote{http://www.mpa-garching.mpg.de/$\sim$swhite/talk/NNG05.pdf}.

The Halo Occupation Distribution (or HOD) method for predicting galaxy
clustering has become popular because it bypasses the need to model
the physics of galaxy formation when analysing the spatial
distribution of galaxies on large scales, for example to constrain the
shape and amplitude of the primordial spectrum of density
fluctuations.  In `classic' HOD models the galaxy population of a halo
depends on its mass alone. This makes it possible to marginalise over
the parameters describing possible occupation distributions in order
to constrain more ``fundamental'' quantities. Many intended
applications require precise measurements and realistic error
estimates, so it is important to quantify any systematic uncertainties
introduced implicitly by the HOD method.  Future large-scale surveys
hope to clarify the nature of Dark Matter and of Dark Energy through
percent-level measurements of the clustering of very large numbers of
galaxies at both low and high redshift (e.g. PanSTARRS
\citep{Kaiser2002} or the Dark Energy Survey \citep{Abbott2005}).
Interpretation will require theoretical models with uncertainties
significantly below this level.

In this paper, we quantify how correlations between halo environment
and halo assembly history affect galaxy clustering.  We use a
simulation of the formation and evolution of the galaxy population
within a very large region ($0.125 h^{-3}{\rm Gpc}^3$) in the
concordance $\Lambda$CDM cosmogony \citep{Croton2006}. This simulation
was carried out by integrating simplified equations for the evolution
of the baryonic component within a stored representation of the
evolving dark matter distribution of the largest high-resolution
cosmological simulation carried out to date, the so-called `Millennium
Simulation' \citep{Springel2005}. Since our galaxy formation modelling
explicitly follows the assembly of each dark matter structure, it
automatically takes care of effects induced by correlations of halo
environment with halo assembly history. We test whether such effects
are significant by randomly swapping the galaxy populations of halos
of identical mass. Such shuffling \emph{does} alter galaxy clustering
on large scales, although it would not if the `classic' HOD assumption
were correct.  Our results should reliably indicate the characteristic
strength of such systematic effects, even if our specific galaxy
formation model is later superseded.

The outline of this paper is as follows.  In Section~\ref{model} we
briefly introduce our simulation and our galaxy formation model.
Section~\ref{method} then describes in detail how we shuffle galaxy
populations between halos of similar properties (i.e. similar mass,
but perhaps also similar concentration or formation redshift). The
differences in clustering between the galaxy distribution in the
original simulation and those produced by such shuffling are explored
as a function of the luminosity and colour of galaxies in
Section~\ref{results}. This quantifies the systematic errors to be
expected in HOD models and addresses the issue of whether they can be
reduced by making the HOD depend on additional halo properties.  We
conclude in Section~\ref{discussion} with a brief discussion and
summary.

\section{Simulation Data}
\label{model}

The Millennium Simulation follows evolution in the distribution of
just over 10 billion dark matter particles in a periodic box of side
$500\,h^{-1}$Mpc. The mass per particle is $8.6\times 10^8\,h^{-1}{\rm
M}_{\odot}$.  The adopted cosmological parameter values are
$\Omega_\Lambda=0.75$, $\Omega_{\rm m}= \Omega_{\rm dm}+\Omega_{\rm
b}=0.25$, $\Omega_{\rm b}=0.045$, $h=0.73$, and $\sigma_8=0.9$,
consistent with a combined analysis of the 2dFGRS \citep{Colless2001}
and first year WMAP data \citep{Spergel2003, Seljak2005}.  The dark
matter distribution is stored at 64 times spaced approximately
logarithmically in expansion factor at early times, and at
approximately 300 Myr intervals after $z\!=\!1$.  Friends-of-friends
(FOF) halos are identified in the simulation at each stored output
with a linking length 0.2 times the mean particle
separation. Substructure is then identified within each halo using an
improved and extended version of the {\small SUBFIND} algorithm of
\citet{Springel2001}.  Having found all halos and their subhalos at
all output times, hierarchical merging trees are constructed which
describe in detail how each structure grows as the universe
evolves. These trees are identical to those used by GSW05 and are the
representation of the evolving dark matter distribution within which
the simulation of galaxy formation is carried out. Further details of
the dark matter simulation and of these procedures can be found in
\cite{Springel2005}.

Our simulation of the formation and evolution of the galaxy population
follows the methodology introduced by \cite{Kauffmann1999} and
extended by \cite{Springel2001}.  Virialised dark matter halos at each
redshift are assumed to have collapsed with their ``fair'' share of
baryons (e.g. $\Omega_{\rm b}/\Omega_{\rm m}$ times their total mass)
from which galaxies form and evolve. The simulation follows the
evolution of the galaxy population in each merger tree. It includes a
wide range of galaxy formation physics using simple, physically based
models tuned to represent both relevant observational data and more
detailed simulations \citep[for detail see][]{Croton2006}.
Importantly, it is the detailed merging, accretion and disruption
histories of the dark matter halos and their substructures that drive
the baryonic modelling and thus ultimately determine the galaxy
content of the halos.  \cite{Croton2006} and \cite{Springel2005} show
that this two-stage simulation scheme can produce a galaxy population
consistent with many observed properties of the local population.
These include the galaxy luminosity function, the bimodal distribution
of colours, the morphology distribution, the Tully-Fisher relation,
and 2-point galaxy correlation functions for samples selected by
luminosity and type.  However, this model is not, of course, perfect.
For example, \cite{Weinmann2006} show that it incorrectly predicts
some aspects of the colour distribution of satellite galaxies in
group-sized halos, and this may impact measurements that are dependent
on colour selection.  To minimise such uncertainties we will always
consider relative measures of bias between shuffled and unshuffled
catalogues to indicate the expected size of the assembly bias effect.
Due to the large volume of the Millennium Simulation our simulated
galaxy catalogue is unprecedented in size, containing $5\,178\,238$
galaxies brighter than $M_{\rm b_J} -\!5\log h = -17$.

\section{The shuffling technique}
\label{method}

If galaxy populations within dark matter halos of a given mass are
statistically independent of all halo properties other than mass, as
assumed in the simplified clustering models described in the
Introduction, then galaxy clustering should not depend on how the
individual realisations of the satellite--central galaxy population
are distributed among the various halos of that mass. We test this by
comparing galaxy correlation functions estimated from the catalogue
described in Section~\ref{model} with identically defined correlation
functions estimated from ``shuffled'' catalogues in which
satellite--central galaxy populations are randomly exchanged between
halos of similar mass.  If halo assembly history is indeed independent
of halo environment, such shuffling should have no effect on the
estimated correlations.

More specifically, for each FOF dark matter halo we first record the
position off-sets of all its galaxies with respect to the `central'
galaxy. This central galaxy sits at the bottom of the halo's potential
well, while further galaxies are satellites which may or may not be
associated with subhalos catalogued by {\small SUBFIND}.  We then
rank-order all halos by virial mass and divide them into mass bins of
width $\log \Delta {\rm M}_{\rm vir} = 0.1$ (although because of the
rapidly decreasing number of halos in the tail of the mass function
the two most massive bins are widened to $\log\, {\rm M}_{\rm vir} =
14.8-15.0$ and $\log\, {\rm M}_{\rm vir} = 15.0-15.5$. Note that the
gradient of the assembly bias effect across this mass range is small,
as shown in Figure~3 of \cite{Wechsler2006}. Note also that here and
elsewhere we define $M_{\rm vir}$ as the mass within the largest
sphere surrounding the halo's potential minimum with mean enclosed
density at least 200 times the critical value).  We then randomly
shuffle the galaxy populations of the halos in each bin.  When doing
this, we take the new central galaxy of each halo to have the same
position as the original central galaxy and we determine the positions
of the new satellites using their recorded off-sets from their central
galaxy. Each central galaxy thus moves together with its own set of
satellites. In the language of HOD modellers
\citep[e.g.][]{Cooray2002} this procedure \emph{exactly} preserves all
1-halo contributions to galaxy clustering statistics. Any differences
between the original and the shuffled catalogues can arise from 2-halo
terms only.  The shuffling is done 10 times with different random
seeds to create 10 different shuffled galaxy catalogues.  It can also
be carried out among halos for which a second variable, such as halo
formation redshift, has been matched in addition to halo mass.  We
show the effect of such extra constraints in Section~\ref{results2}.

To quantify the difference in clustering between the actual and the
shuffled galaxy catalogues we measure the 2-point autocorrelation
function for each and plot their relative bias, $b(r)$, defined by
\begin{equation}
b(r) = \Big( \frac{\xi_{\rm orig}}{\xi_{\rm shuff}} \Big)^{1/2}~.
\label{eq1}
\end{equation}
Here $\xi_{\rm shuff}$ is the 2-point function of the shuffled
catalogue at pair separation $r$, and $\xi_{\rm orig}$ is the
corresponding 2-point function for the original (unshuffled)
catalogue. Note that a value of $b\!>\!1$ implies that the shuffling
dilutes the clustering of the original distribution.  Note also that
whenever we estimate $b$ below, exactly the same galaxy set is use to
estimate both $\xi_{\rm orig}$ and $\xi_{\rm shuff}$.  Only the
positions of the galaxies are changed by the shuffling.

\section{Results}
\label{results}

\subsection{The strength of second parameter effects}
\label{results0}

\begin{figure}
\plotone{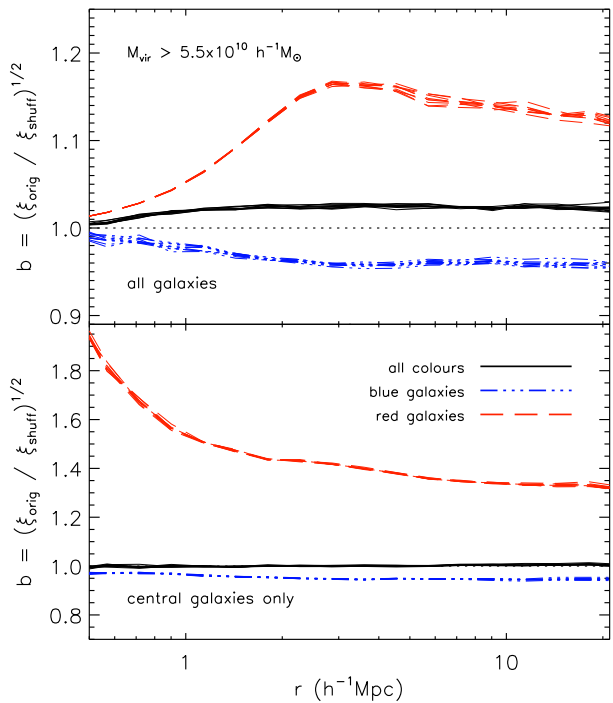}
\caption{The relative bias between the original and the shuffled
galaxy populations in subhalos more massive than $5.5\times 10^{10}
h^{-1} {\rm M}_{\odot}$ (i.e. $>64$ simulation particles) as a
function of pair separation (Eq.~\ref{eq1}).  The top panel shows
results for all galaxies, whereas the bottom panel is restricted to
central galaxies (resulting in one and only one galaxy per halo).  In
each panel solid lines refer to the full sample, while long-dashed
lines are for blue galaxies and dashed-dotted lines for red galaxies.
The two subpopulations are split at ${\rm B-V}=0.8$. Strong bias
effects are seen in a number of cases demonstrating that the galaxy
content of a halo of given mass is correlated with the halo's
large-scale environment.  }
\label{fig1}
\end{figure}

In Fig.~\ref{fig1} we plot the relative bias between our 10 shuffled
galaxy catalogues and the original Millennium Run catalogue as a
function of pair separation, and for subsets of galaxies selected in
various ways.  In this subsection we show results for subcatalogues
which contain only galaxies in subhalos with mass \citep[as defined by
{\small SUBFIND}, see][]{Springel2005} greater than $M_{\rm
vir}\!=\!5.5\times 10^{10} h^{-1} {\rm M}_{\odot}$ (i.e. $>64$
simulation particles).  This means that we consider only galaxies
which reside in well-resolved dark matter (sub)structures at $z=0$. In
the top panel, relative bias functions are shown for this sample as a
whole and for subsamples split by colour at ${\rm B\!-\!V}=0.8$
\citep[see Fig.~9 of][]{Croton2006}.  The bottom panel presents a
similar analysis but further restricts the catalogues to contain only
the central galaxies of the halos.  Note that for all statistics there
is a very small scatter between the 10 relative bias measurements.
This demonstrates that `small sample' effects are negligible for the
questions we address here.

Consider first the top panel of Fig.~\ref{fig1}. The galaxy population
as a whole (the solid lines) shows a systematic bias of $\sim\!3\%$ on
large scales. Shuffling has reduced the strength of clustering by a
small but significant amount.  Note that this result is independent of
the galaxy formation model, since shuffling does not change the set of
central galaxy positions but merely reassigns populations of
well-resolved subhalos among halos of similar mass. Clearly the
assembly histories of dark halos are \emph{not} independent of their
clustering properties (as GSW05 already showed) and this \emph{does}
affect galaxy clustering\footnote{A preliminary k-space analysis by
Nikhil Padmanabhan using one of our shuffled catalogues suggests that
the effects of assembly bias on the power spectrum amplitude are
non-negligible out to scales of at least $k\!\sim 0.05 h\,{\rm
Mpc}^{-1}$ ($>\!100 h^{-1}{\rm Mpc}$), after which noise dominates the
signal.  We leave a detailed power spectrum analysis of assembly bias
for future work.}.  Red galaxies (long-dashed lines) are biased in the
same way as the sample as a whole but at the $\sim\!15\%$ level, while
blue galaxies (dashed-dotted lines) are biased with the opposite sign
at the $\sim\!5\%$ level.  These results do, of course, depend on the
galaxy formation model which determines whether galaxies are red or
blue.  The overall bias is effectively a weighted average of these two
partially compensating effects.  Note that bias is negligible on small
scales and grows to a value which is almost constant for $r \simgt 3
h^{-1}{\rm Mpc}$. This reflects the fact that only the 2-halo term
contributes (i.e. clustering between galaxies which live in
\emph{different} halos). This is diluted on small scales by 1-halo
clustering which is identical in all catalogues.  The change in
clustering amplitude for galaxies (here $(1\!-\!b)^2 \simlt 30\%$) is
smaller than that found by GSW05 for dark matter halos (which was up
to a factor of $\sim\!5$).  This is because we sum clustering
contributions from halos with a wide range in mass, thereby diluting
the predominantly low-mass GSW05 effect.

When we consider the clustering of central galaxies only the total
number of galaxies in these catalogues is reduced by approximately
30\% and the relative bias functions change considerably.  By
definition, there is now one and only one galaxy in each dark halo so
there is no 1-halo contribution to the correlation functions. In
addition, the correlation function for the population as a whole is
invariant under shuffling.  Thus the solid lines in the lower panel of
Fig.~\ref{fig1} all coincide with $b(r)=1$. There are, however,
substantial effects when the population is split by colour,
demonstrating that the colour of the central galaxy in a halo of given
mass depends significantly on the halo's environment. Halos with red
central galaxies show a strong relative bias ($\sim\!40\%$ on large
scales, rising to $\sim\!80\%$ on small scales) while halos with blue
central galaxies show a weaker one which is very similar to that for
all blue galaxies ($\sim\!5\%$).  The strong effect for red central
galaxies reflects the fact that such objects are found primarily in
two very specific types of halo: massive halos where cooling and
star-formation have been curtailed by AGN feedback; and lower mass
halos which have just passed through a more massive system, thereby
losing their hot gas atmospheres and so their source of fuel for star
formation. Both cases are associated with a massive halo, hence the
high clustering amplitude. The great majority of central galaxies are
associated with more isolated and/or lower mass halos and have ongoing
star formation; these objects are blue.

\subsection{Assembly bias as a function of galaxy luminosity}
\label{results1}

\begin{figure}
\plotone{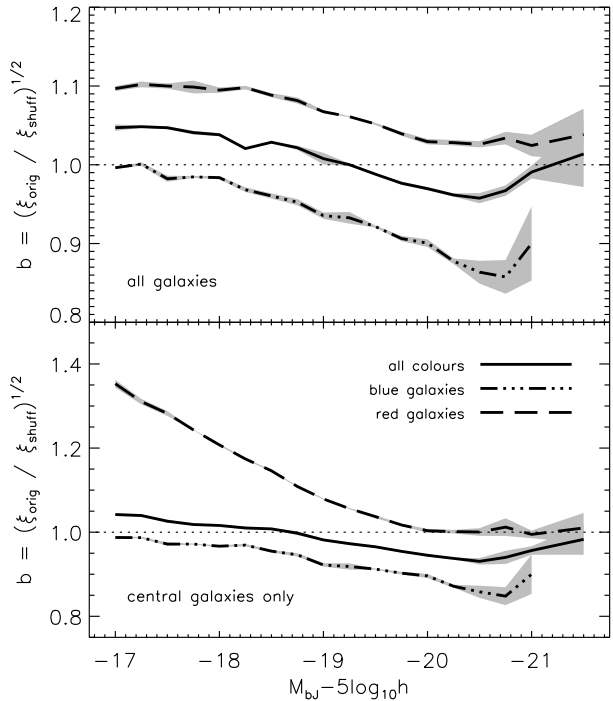}
\caption{The assembly bias (i.e. the enhancement in clustering induced
by correlations between halo assembly history and large-scale
environment at fixed halo mass) as a function of magnitude limit for
absolute magnitude limited samples of galaxies. The upper panel gives
results for all galaxies and the lower panel for central galaxies
only. In each panel the solid line gives results for galaxies of all
colours, while the long-dashed and dashed-dotted lines are for blue
and red subsamples respectively. The two samples are split at ${\rm
B-V}=0.8$.  The grey shaded region surrounding each line indicates the
$1\sigma$ scatter in assembly bias values for our 10 shuffled
catalogues.  Significant bias (i.e. $b\!\ne\!1$) is seen for almost
all galaxy subsamples.}
\label{fig2}
\end{figure}

We now generalise the above results including all galaxies which are
well resolved by the formation model regardless of their subhalo mass
at $z\!=\!0$. Fig.~\ref{fig2} shows the relative bias between the
shuffled and the original galaxy populations as a function of both
colour \emph{and} luminosity. On scales $r \simgt 3h^{-1}{\rm Mpc}$
1-halo terms do not contribute to the correlations and the relative
bias is approximately constant for all samples we have considered.
For simplicity we therefore average the relative bias measurements for
each of our 10 shuffled catalogues over the separation range
$6\!-\!12h^{-1}{\rm Mpc}$ and we characterise the result by the mean
and $1\sigma$ scatter of these values.  In the following we refer to
this quantity as the {\it assembly bias} as it measures the bias
induced by the environmental dependence of halo assembly history at
fixed halo mass.

The top panel of Fig.~\ref{fig2} shows this assembly bias for absolute
magnitude limited subsamples of galaxies as a function of their
magnitude limit.  Again we plot results for galaxies of all colours
(solid line) and for blue (dot-dashed line) and red (long-dashed line)
galaxies separately.  The bottom panel shows an identical analysis but
for samples restricted to central galaxies.  Note that selecting
galaxy subsamples by limiting stellar mass rather than luminosity
produces similar behaviour to that presented below.  This is expected
given that the scatter in $\log({\rm M}/\rm{L})$ for the galaxies is
typically small in comparison with the magnitude range over which
the assembly bias changes.

If we focus first on the upper panel of Fig.~\ref{fig2}, we see that
correlations between assembly history and environment at fixed halo
mass can either enhance (for faint galaxies) or dilute (for bright
galaxies) the strength of galaxy clustering, with a transition near
the characteristic luminosity $L_*$ of the galaxy luminosity function.
Fainter than $M_{\rm b_J}\!-5\log h \sim -20.5$ bias values for the
red and blue subpopulations are symmetrically offset from the curve
for the population as a whole by about 5\%.  Brighter than this, the
bias for the population as a whole approaches that for the red
subpopulation, reflecting the fact that there are few blue galaxies at
these magnitudes. At $M_{\rm b_J} \!-5\log h \sim -20$ blue galaxies
have an assembly bias of about 0.9, showing that they occupy halos
with significantly lower density environments than randomly selected
halos of the same mass.

In the bottom panel of Fig.~\ref{fig2} we show the assembly bias for
absolute magnitude limited samples of {\it central} galaxies (i.e. for
samples of halos defined by the luminosity and colour of their central
galaxies).  A notable difference from the central galaxy samples
studied in the bottom panel of Fig.~\ref{fig1} (which were defined by
the {\it mass} of their halos) is that the assembly bias differs from
unity not only for the red and blue subsamples but also for samples
without colour selection.  This difference is caused by scatter in the
relation between halo mass and central galaxy luminosity which
correlates with halo environment in a way that is different for halos
with faint ($L\!<\!L_*$) and with bright ($L\!>\!L_*$) central
galaxies.  Low-mass halos with brighter than average central galaxies
are in denser than average environments, while the opposite is true
for higher mass halos.  At all magnitudes blue central galaxies
inhabit halos with lower density environments than red central
galaxies. This is in part because at given absolute magnitude blue
central galaxies tend to have lower mass halos than red ones.

From Fig.~\ref{fig2} we see that assembly bias is strongest for faint
red central galaxies.  These galaxies reside at the centres of
low-mass ($\sim\!10^{11}M_{\odot}$) dark matter halos and have a $b$
value of about 1.4, which translates to an autocorrelation amplitude
about twice that which would have been found if their halos had been
randomly chosen according to their mass alone.  The mean formation
redshift of halos with $-17\!>\!M_{\rm b_J}\!-5\log h \!>\!-18$ red
central galaxies is $z_{\rm form}\!=\!2.9$, whereas blue central
galaxies of similar magnitude have halos with $z_{\rm
form}\!=\!1.8$. As noted above, many of the faint red central galaxies
occupy halos which have recently passed through a much more massive
system. As a result they have both high formation redshifts and high
density environments. This accounts for much of their strong assembly
bias. The bias for {\it all} faint red galaxies (see the top panel) is
less pronounced due to dilution by satellites in group and cluster
mass halos. As GSW05 showed, the correlation between assembly history
and environment is much weaker for these than for low-mass halos. In
contrast, the blue galaxy curves are similar in the top and bottom
panels: this is simply because most blue galaxies are central
galaxies.  \emph{Bright} red central galaxies are largely unaffected
by assembly bias because they have high mass halos. Such halos almost
never have blue central galaxies and for them the GSW05 effect is, in
any case, weak.

The extent to which assembly bias effects are due to the properties of
satellite galaxies rather than to those of central galaxies can be
tested by shuffling satellite populations as before while keeping all
central galaxies in their original positions. We have done this,
finding relatively weak effects.  If we shuffle satellites only, we
find an assembly bias $(\xi_{\rm orig}/\xi_{\rm shuff})^{1/2} \approx
1.02$ to all magnitude limits for the full galaxy population.  For the
red galaxies, $(\xi_{\rm orig}/\xi_{\rm shuff})^{1/2} \approx 1.04$,
while for the blue galaxies there is no significant effect, $(\xi_{\rm
orig}/\xi_{\rm shuff})^{1/2} \approx 1.0$.  The fact that shuffling
satellites alone changes the correlation amplitude suggests that the
satellite population of a halo somehow ``knows'' about its large-scale
environment, in the sense that halos with many neighbours tend to have
more substructure than similar mass halos with few. A weak tendency
for halos with neighbours to have above-average substructure was
already noted by \cite{Wechsler2006}.

\subsection{A second variable?}
\label{results2}

\begin{figure}
\plotone{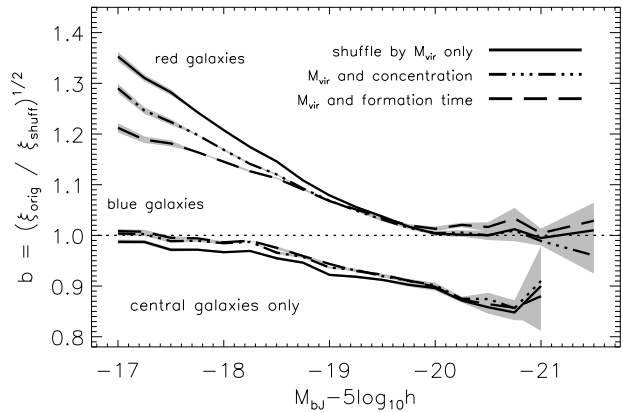}
\caption{Assembly bias for absolute magnitude limited samples of red
and blue central galaxies (i.e. for halos with red or blue central
galaxies brighter than a specified limit).  Results are shown for
three different implementations of the shuffling procedure of
Section~\ref{results2}: swapping among halos of similar virial mass
(replotted from Fig.~\ref{fig2}, solid lines); swapping among halos of
similar virial mass {\it and} similar concentration (dashed-dotted
lines); and swapping among halos of similar virial mass {\it and}
similar formation time (long-dashed lines).  The grey shaded region
surrounding each line indicates the $1\sigma$ scatter among 10
shuffled catalogues. Assembly bias is somewhat weaker when halo
concentration or formation time is specified in addition to mass, but
it is not eliminated.}
\label{fig3}
\end{figure}

\begin{figure*}
\plotfull{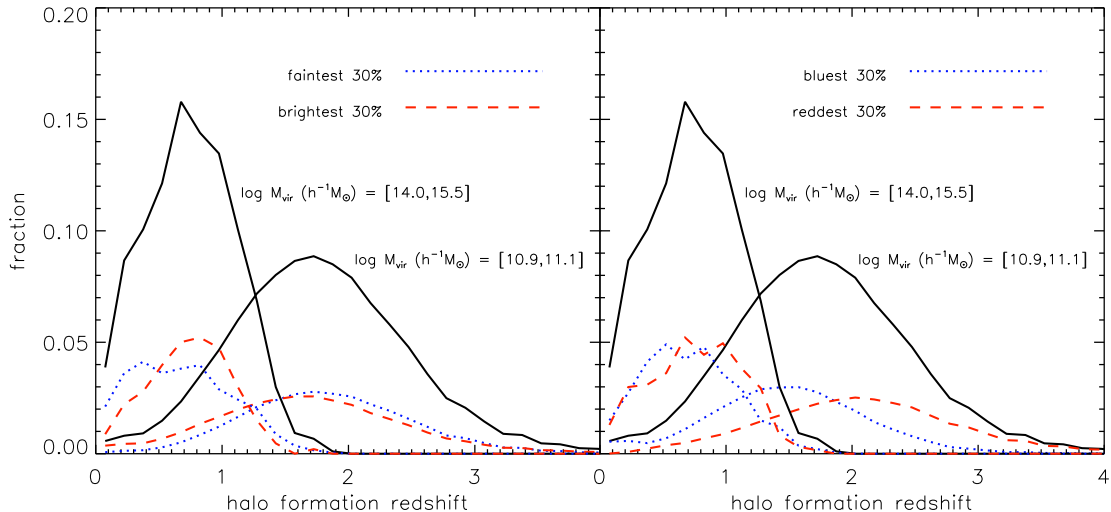}
\caption{The distribution of halo formation redshifts for two ranges
of halo mass as indicated. Solid lines show the distribution for all
halos in each mass range and are the same in both panels. Dotted lines
in the left panel show distributions for the $30\%$ of halos in each
mass range with the faintest central galaxies, while dashed lines are
for the $30\%$ tail with the brightest central galaxies. Dotted and
dashed lines in the right panel are similar except they now refer to
the halos with the bluest and reddest central galaxies respectively.
For lower mass halos the colour, and to some extent also the
luminosity, of the central galaxy is correlated with formation
redshift. At high masses such effects may also be present but our
results are noisier because of the much smaller number of halos
involved.}
\label{fig4}
\end{figure*}

The above results show that populating dark halos with a semi-analytic
or HOD algorithm based on halo mass alone will result in correlation
functions which have systematic errors between 5\% and a factor of 2
depending subsample definition.  We now ask whether more complex
algorithms which include dependences on additional halo properties can
account for the assembly bias in our simulated catalogues. At some
level this must be possible since the galaxy content of each simulated
halo is {\it determined} by its detailed assembly history. It is
unclear, however, whether this history is suitably summarised by
parameters such as formation redshift or concentration.
(\cite{Navarro1996}, \cite{Wechsler2002} and \cite{Gao2004} have shown
that both these properties are closely related to the growth history
of a halo's main progenitor.)  To explore this issue we consider two
more highly constrained shuffling procedures: swapping galaxy
populations between halos of similar mass \emph{and} similar formation
redshift, and between halos of similar mass \emph{and} similar
concentration.  In the following, formation redshift is defined as the
redshift when a halo's main progenitor contains half its final mass,
as used by \cite{Gao2004} and GSW05. We linearly interpolate halo
growth between simulation outputs to increase the precision with which
this redshift can be estimated.  To estimate halo concentration we
take the measured $V_{\rm vir}$ and $V_{\rm max}$ for each halo and
solve Eq.~5 in \cite{Navarro1996}.

In Fig.~\ref{fig3} we show the result of this exercise. We plot
relative bias as defined above (again estimated from an ensemble of 10
shuffled catalogues) for absolute magnitude limited central galaxy
subsamples split by colour. These were the subsamples with the most
pronounced effects in Fig.~\ref{fig2} and the results from that figure
are repeated here as solid lines.  The other lines show how the
relative bias is reduced when shuffling preserves halo formation
redshift or halo concentration in addition to halo mass, thus how much
of the assembly bias in the original simulated catalogue can be
represented using these additional halo properties.  Note that results
for the ``all colour'' and ``all galaxy'' catalogues presented in
Figure 2 show dependences on these additional parameters that are
much weaker than the extreme cases shown here, typically less than
a few percent.  We thus omit them for clarity.

Interestingly, Fig.~\ref{fig3} shows that neither formation redshift
nor concentration encodes sufficient information to account for the
assembly bias of the simulated galaxy catalogue.  Of the two
parameters, formation redshift is the most successful, accounting for
about 40\% of the assembly bias for faint red central galaxies (at
$M_{\rm b_J}\!-5\log h = -17$ the relative bias is reduced from 1.37
to 1.22) but only a few percent of the assembly bias for bright blue
central galaxies. Employing concentration as the second parameter is
only about half as effective in reducing the relative
bias. Concentration dependences can account for only a small fraction
of the measured assembly bias. Clearly, although the galaxy content of
our simulated halos depends only on their mass and their assembly
history, there is some aspect of the assembly history which is not
encoded in halo concentration or formation redshift and which
correlates with large-scale environment.  Fig.~\ref{fig3} demonstrates
that halo concentration and halo formation redshift do not encode the
same information about large-scale clustering and that neither
provides the information needed for precise modelling of the
large-scale clustering of galaxies.

Fig.~\ref{fig4} explores the relation of central galaxy properties to
formation redshift in more detail. It shows the distribution of
formation redshift for dark matter halos in two mass ranges, $\log
M_{\rm vir}\!>\!14.0$ and $\log M_{\rm vir}\!=\!10.9-11.1$ $(h^{-1}
M_\odot)$.  In each range we compare the distribution for all halos
(solid lines) with those for subpopulations which are extreme in their
central galaxy properties, either luminosity or colour. Dashed and
dotted curves in the left panel correspond to the 30\% tails
containing the brightest and the faintest central galaxies
respectively, while in the right panel they refer instead to the 30\%
bluest and reddest central galaxies.

Fig.~\ref{fig4} shows the well known result that low-mass halos have a
broad distribution of formation redshifts, centred at $z\!\sim\!1.7$
and extending from $z\!>\!3$ to $z\!<\!0.5$, while high-mass halos
formed more recently, with a distribution centred at $z\!\sim\!0.7$
and with tails extending from $z\!\sim\!0.0$ to $z\!\sim\!1.5$
\citep[see][]{Lacey1993}.  For low-mass halos there is a weak but
significant shift in formation redshift distribution between those
with bright and those with faint central galaxies: faint central
galaxies live in halos that formed systematically earlier than those
of their brighter counterparts.  (In this mass bin the mean absolute
magnitudes of the $30\%$ faintest and 30\% brightest central galaxies
are $-17.2$ and $-17.9$, respectively).  The converse appears true in
high-mass halos: brighter central galaxies tend to occupy halos that
formed earlier, while fainter central galaxies occupy halos that
formed later.  (The mean absolute magnitudes of the $30\%$ faintest
and brightest central cluster galaxies are $-21.1$ and $-22.1$
respectively).

For low-mass halos the effects as a function of colour are
significantly stronger. The right panel of Fig.\ref{fig4} shows that
halos with red central galaxies have significantly earlier formation
redshifts than those with blue.  The mean ${\rm B\!-\!V}$ colour of
the central galaxy shifts from $0.44$ to $0.69$ between the two tails.
For high-mass halos there is little difference in the formation
redshift distribution between those with the reddest and those with
the bluest central galaxies. This is simply because most of the
central galaxies in high-mass halos are red; the shift in colour
between the two tails is only from $0.90$ to $0.94$ in this case.  In
combination with the effect discovered by GSW05, the correlation
between central galaxy colour and halo formation redshift in low-mass
halos explains a significant part (roughly half) of the large-scale
assembly bias which we measure for faint red central galaxies (see
Fig.~\ref{fig3}).

\section{Discussion and conclusions}
\label{discussion}

We ask a simple question in this paper: to what degree is galaxy
clustering influenced by the assembly bias of dark halos, the fact
that their clustering depends not only on their mass but also on the
details of their assembly history?  Such dependences are neglected in
the halo occupation distribution schemes that have become popular for
constructing galaxy catalogues from dark matter simulations. Our
results show that they are significant, however, and thus may
introduce systematic errors if HOD techniques are used to derive
cosmological parameters from the clustering in large galaxy surveys.
It appears that this problem is not easily addressed by including an
additional halo parameter in HOD models. Detailed tracking of galaxy
formation during halo assembly seems necessary.

Our conclusions are based on analysis of galaxy clustering in a very
large simulation in which the formation of galaxies has been followed
explicitly during halo assembly. By comparing galaxy catalogues drawn
from this simulation with `shuffled' catalogues where galaxy
populations have been swapped among halos of similar properties, we
can measure the sensitivity of galaxy clustering to the details of
halo assembly.  Our principal results can be summarised as follows:

\begin{itemize}

\item Assembly bias can be significant and can be of either sign.  The
effects differ qualitatively between galaxy samples selected above a
halo mass limit and those selected above a galaxy luminosity limit, as
well as between samples containing all galaxy types and those
containing only the central galaxies of halos.  In addition they
depend on galaxy colour.  The strongest effects are found for
low-luminosity, red central galaxies. Assembly bias enhances the
amplitude of their 2-point correlations by almost a factor of 2.

\item Simulation galaxies selected to a faint absolute magnitude limit
(e.g. $M_{\rm b_J}\!-5\log h < -17$) are more strongly clustered than
they would be if halos of a given mass had galaxy populations
distributed independently of other halo properties. This effect
reverses for samples selected above a relatively bright absolute
magnitude limit (e.g. $M_{\rm b_J}\!-5\log h < -20$). In both cases the
bias alters the amplitude of the 2-point correlation function of the
galaxies by up to 10\%.

\item When absolute magnitude limited galaxy samples are split by
colour at ${\rm B\!-\!V} = 0.8$, the blue and red subsamples have values
of assembly bias which are off-set from the value for their parent
samples by $-5\%$ and $+5\%$ respectively, corresponding to 10\%
off-sets in autocorrelation amplitude.

\item As expected from the results of GSW05, halo formation redshift
encodes some of the effects leading to assembly bias. Surprisingly,
however, allowing the galaxy populations of halos to depend on halo
formation redshift in addition to halo mass accounts for only 40\% of
the assembly bias for red central galaxies and has no influence on
that for blue central galaxies. Halo concentration is even less
successful as a second parameter, only half as effective as formation
redshift in accounting for assembly bias in the simulation. Most of
this bias must be due to a correlation between other aspects of halo
assembly and halo environment.

\item As is well known, dark matter halos of a given mass show a wide
range of formation redshifts.  For given mass, the distribution of
formation redshift depends both on the colour and on the luminosity of
the central galaxy. The effects are in most cases quite weak, however,
reinforcing the impression that other aspects of halo formation must
be responsible for the strong assembly bias we find for colour- and
absolute magnitude selected samples of central galaxies.

\end{itemize}

New large-scale galaxy surveys, such as PanSTARRS \citep{Kaiser2002}
and the Dark Energy Survey \citep{Abbott2005}, are currently being
designed to obtain extremely precise measures of galaxy clustering at
a variety of redshifts. The goal is to use these to infer the linear
power spectrum of density fluctuations, the rate at which it grows
with redshift, and the recent expansion history of the Universe.
These quantities then constrain the nature of Dark Matter, the nature
of Dark Energy, and the process which created all cosmic structure.
Significant conclusions will require measures of galaxy clustering to
be translated into estimates of more fundamental cosmological
parameters (e.g. the amplitude of linear fluctuations at each
redshift, the characteristic scale of baryon wiggles, the effective
slope of the primordial power spectrum...) with a precision of a few
percent or better.  The stated goal for the HOD machinery is to
convert observed clustering measures to fundamental quantities at this
level of precision, while bypassing the need to understand the details
of galaxy formation. Our results suggest that the details of galaxy
formation \emph{do} affect clustering statistics at at least the 5\%
level in a way which cannot easily be included in an HOD model.

It is important to stress that we do not claim that our galaxy
formation model is correct, just that it is plausible, and so can be
used to explore the size of assembly bias effects. For many
applications a systematic error in galaxy correlation amplitudes at
the $5$ or $10$ percent level can safely be ignored.  Nevertheless, if
we wish to understand galaxy formation we must clearly model it. The
results presented in this paper show not only that galaxy formation
influences large-scale clustering in unexpected ways which are not
consistent with current simplified clustering models, but also that
these models may be subject to systematic errors which make them
unsuitable for interpreting precision measures of galaxy clustering in
terms of fundamental physics. A deeper understanding of galaxy
formation appears required to carry through this programme
successfully.

\section*{Acknowledgements}
\label{acknowledgements}

The authors would like to thank Nikhil Padmanabhan for measuring the
Fourier space assembly bias using one of our shuffled catalogues.
Special thanks to the referee, David Weinberg, for valuable comments
that improved the quality of this paper. DC acknowledges support from
NSF grant AST00-71048.  The Millennium Run simulation used in this
paper was carried out by the Virgo Supercomputing Consortium at the
Computing Centre of the Max-Planck Society in Garching.  The galaxy
catalogues used here are publicly available at
http://www.mpa-garching.mpg.de/galform/agnpaper.

\bibliographystyle{mnras}
\bibliography{./paper}

\label{lastpage}

\end{document}